\documentclass[aps,prl,twocolumn,superscriptaddress,longbibliography]{revtex4-2}
\usepackage{graphics}
\usepackage{graphicx}
\usepackage{bbm,bm}
\usepackage{xcolor}

\ifx\pdftexversion\undefined
\usepackage[dvips]{hyperref}
\else
\usepackage{hyperref}
\fi
\hypersetup{
  colorlinks = true, linkcolor = blue
}

\usepackage[normalem]{ulem}

\begin{document}

\title{Observation of $1/k^4$-tails after expansion of Bose-Einstein Condensates with impurities
}

\author{Hugo Cayla}
\affiliation{Universit\'e Paris-Saclay, Institut d'Optique Graduate School, CNRS, Laboratoire Charles Fabry, 91127, Palaiseau, France}
\author{Pietro Massignan}
\affiliation{Departament de F\'isica, Universitat Polit\`ecnica de Catalunya, Campus Nord B4-B5, E-08034 Barcelona, Spain}
\author{Thierry Giamarchi}
\affiliation{Department of Quantum Matter Physics, University of Geneva,
24 quai Ernest-Ansermet, 1211 Geneva, Switzerland}
\author{Alain Aspect} 
\affiliation{Universit\'e Paris-Saclay, Institut d'Optique Graduate School, CNRS, Laboratoire Charles Fabry, 91127, Palaiseau, France}
\author{Christoph I. Westbrook}
\affiliation{Universit\'e Paris-Saclay, Institut d'Optique Graduate School, CNRS, Laboratoire Charles Fabry, 91127, Palaiseau, France}
\author{David Cl\'ement}
\affiliation{Universit\'e Paris-Saclay, Institut d'Optique Graduate School, CNRS, Laboratoire Charles Fabry, 91127, Palaiseau, France}

\date{\today}

\begin{abstract}
We measure the momentum density in a Bose-Einstein condensate (BEC) with dilute spin impurities after an expansion in the presence of interactions. We observe tails decaying as $1/k^4$ at large momentum $k$ in the condensate and in the impurity cloud. These algebraic tails originate from the impurity-BEC interaction, but their amplitudes greatly exceed those expected from two-body contact interactions at equilibrium in the trap. Furthermore, in the absence of impurities, such algebraic tails are not found in the BEC density measured after the interaction-driven expansion. These results highlight the key role played by impurities when present, a possibility that had not been considered in our previous work [Phys. Rev. Lett. {\bf 117}, 235303 (2016)]. Our measurements suggest that these unexpected algebraic tails originate from the non-trivial dynamics of the expansion in the presence of impurity-bath interactions. 
\end{abstract}

\maketitle 

Impurities strongly affect the properties of low-temperature ensembles of particles. Well-known examples range from the Kondo effect \cite{goldhaber1998, cronenwett1998, liang2002, alloul2009}, to quantum localization \cite{balatsky2006,  peres2010, ospelkaus2006} and polar crystals \cite{frohlich1954, feynman1955}. Similarly, the transport of massive impurities strongly depends on the medium in which they propagate and scatter, as illustrated in experiments conducted with gaseous Bose-Einstein condensates (BECs) as a bath \cite{chikkatur2000, palzer2009,zipkes2010, schmid2010, catani2012,fukuhara2013, balewski2013}. Impurities can also serve as accurate probes for equilibrium and out-of-equilbrium properties of their many-body environment \cite{daley2004, klein2005, ng2008, hohmann2015, skou2021}. 

Theoretical approaches introduce a quasi-particle, the polaron, which describes the dressing of the impurity by the collective excitations of the bath \cite{landau1933,pekar1946}. In quantum gas experiments, such descriptions were validated in ensembles of both bosons and fermions \cite{devreese2009, massignan2014,scazza2022}. Furthermore, the capability to tune interactions in atomic gases permitted detailed studies beyond the weakly-interacting regime \cite{jorgensen2016, hu2016, yan2020,devreese2020}.
However, recent theoretical work \cite{massignan2021, schmidt2021, levinsen2021} illustrated the need for a precise knowledge of the impurity-bath interaction to accurately describe Bose polarons and quantities such as their energy in the strongly-interacting regime. Tan's contact provides direct information on short-ranged interaction potentials in systems at equilibrium \cite{tan2008} and it was measured recently for Bose polarons \cite{yan2020}. Here, we investigate an out-of-equilibrium configuration and observe unexpected signals resulting from the impurity-BEC interaction.

In our experiments we measure momentum densities after an expansion of dilute BECs containing weakly-interacting impurities. We observe that: {\it (i)} the momentum density of the impurities measured after an expansion exhibits algebraic tails whose decay is consistent with $1/k^4$ over a surprisingly large momentum range, {\it (ii)} the algebraic tails are observed only in the presence of both the BEC and the impurities, {\it (iii)} the tail amplitude increases linearly with the number of impurities, and {\it (iv)} no such tails are identified when the impurities are immersed in a thermal gas (instead of a BEC). These features qualitatively resemble those expected from two-body interactions at equilibrium in the trap. However, the tails amplitude is orders of magnitudes larger than the one expected from the in-trap impurity contact. In fact, our observation is even more surprising, because the in-trap $1/k^4$-tails associated with the impurity-BEC contact are expected to vanish adiabatically during the expansion in the presence of interactions \cite{qu2016}. This suggests that the expansion dynamics of the impurity-bath system plays a key role in our observations.

We also revisit our previous observation of $1/k^4$-tails in an expanding BEC \cite{chang2016}. Our earlier experiment was conducted under conditions identical to the present ones. Dilute impurities were very likely present there as well, but we were not aware of their presence. Here we show unambiguously that the tails disappear when the impurities are removed from the BEC, and therefore we confirm experimentally the scenario predicted theoretically  \cite{qu2016}.

\begin{figure}[ht!]
\includegraphics[width=\columnwidth]{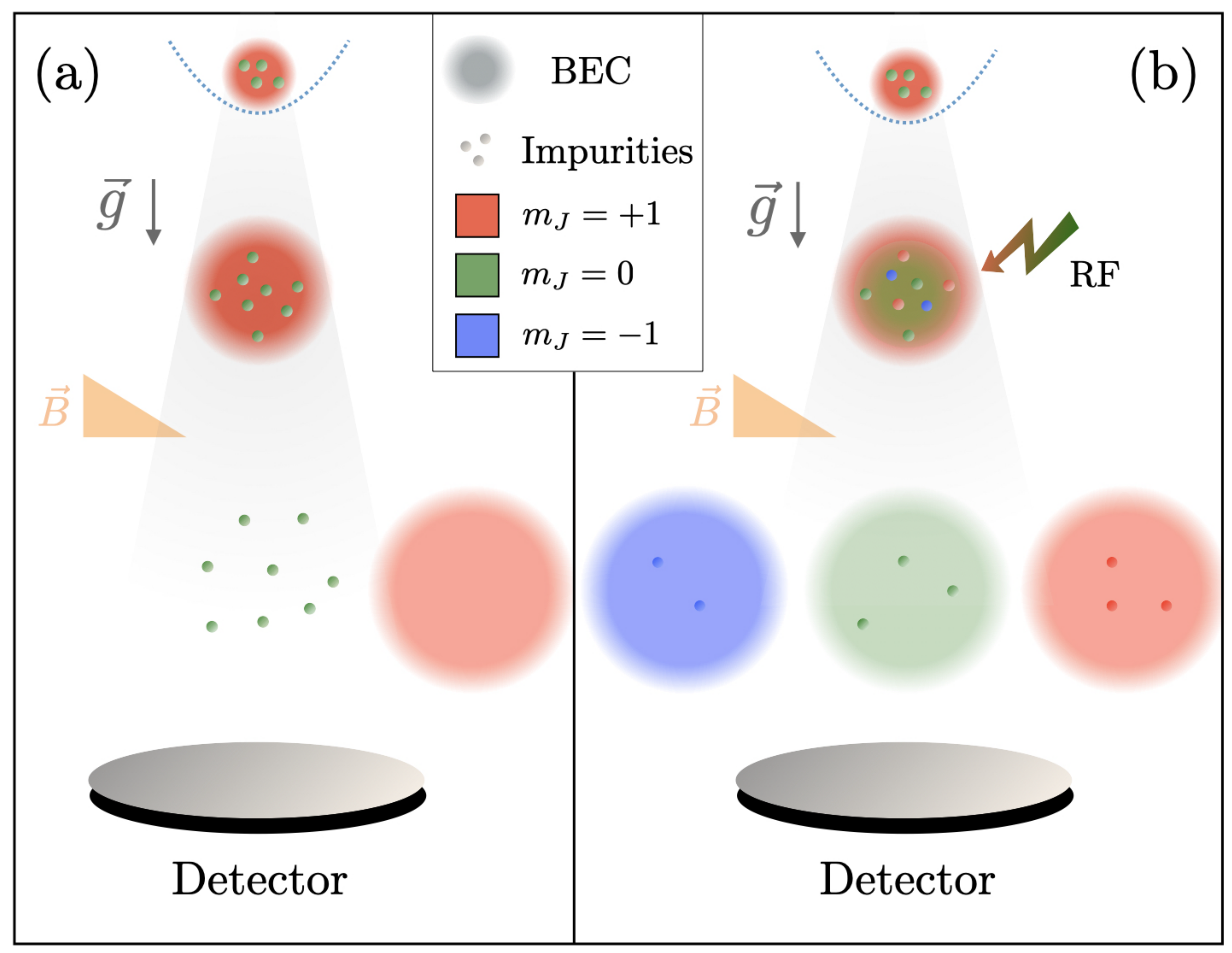}
\caption{
{\bf (a)} To detect only spin impurities ($m_{J}=0$), a magnetic gradient is applied during the expansion to push the BEC atoms ($m_{J}=+1$) away from the detector. {\bf (b)} To probe simultaneously BEC atoms and impurities, a radio-frequency (RF) pulse couples the atomic states $m_{J}=0$ and $m_{J}=+1$ during the expansion, before applying the magnetic gradient, producing a known mixture of $m_{J}=0$ and $m_{J}=\pm 1$.
}
\label{fig1}
\end{figure}

In our experiment the bath is a BEC of metastable Helium-4 ($^4$He$^*$) atoms in the $m_{J}=+1$ sub-level and the impurities are a small number of $^4$He$^*$ atoms in the $m_{J}=0$ sub-level. The impurity-bath scattering length $a_{IB}\simeq142~a_{0}$ is equal to the scattering length within the BEC, $a_{BB}\simeq 142~a_{0}$ \cite{leo2001}, with $a_{0}$ the Bohr radius. 
Our experiment starts with the production of a degenerate Bose gas of $^4$He$^*$ atoms in a crossed optical dipole trap (ODT) \cite{bouton2015}. 
The evaporation to BEC is performed with (most of) the $^4$He$^*$ atoms polarised in the $m_{J}=+1$ magnetic sub-level to avoid the large inelastic collisions present in non-polarised $^4$He$^*$ gases \cite{shlyapnikov1994}. The ODT frequencies at the end of the evaporation are $\omega_{x}/2\pi=110$ Hz, $\omega_{y}/2\pi=400$ Hz and $\omega_{z}/2\pi=420$ Hz. The polarisation of the atoms is maintained in the optical trap with a magnetic bias field of $\sim4~$G, but this does not ensure a full spin-polarisation of the gas. Indeed, we recently discovered that a small fraction ($f_{I}\lesssim 1\%$) of spin impurities ($m_{J}=0$ atoms) is present in the BEC. These impurities originate from spin flips occurring during the loading of the optical trap from a magnetic quadrupole trap \cite{bouton2015}.  This previously unnoticed situation is an opportunity to study Bose polarons in a particular setting. Indeed, given that $a_{BB}$ and  $a_{IB}$ are equal, the mean-field interaction of impurities with the BEC exactly cancels the harmonic trapping potential. As a consequence, spin impurities are trapped in a perfectly flat potential in the region where the BEC is present \cite{Imp_mJ-1}. 

In the following, we concentrate on the tails of the momentum density at large momenta, exploiting our ability to record extremely small densities of $^4$He$^*$ \cite{cayla2018}. Our investigation builds upon {\it (i)} tuning the number of impurities and the atom number in the BEC independently, and {\it (ii)} detecting selectively the spin sub-levels.  We vary the BEC atom number $N_{\rm BEC}$ from $1 \times 10^4$ to $1 \times 10^6$ by modifying the shape of the optical trap during the evaporation process \cite{ExpDetail}. The impurity fraction $f_{I}$ is varied  between 0.05\% and 1\% using optical pumping and varying holding times in the trap  \cite{ExpDetail}. Once we have prepared the trapped gas in the conditions we are interested in, we switch off the trap and let the gas expand in free-fall for a long time-of-flight (TOF) $t_{\rm TOF}=298 \ $ms. Exploiting the different magnetic properties of  $m_{J}=+1$ and $m_{J}=0$ atoms, we selectively detect the two spin sub-levels. More precisely, we can choose to detect only atoms initially trapped  in the $m_{J}=0$ state by pulsing a magnetic gradient during the expansion to push the $m_{J}=+1$ atoms away from the detector (see Fig.~\ref{fig1}a). Alternatively, we can detect a known fraction of both initially trapped $m_{J}=+1$ and $m_{J}=0$ atoms (see Fig.~\ref{fig1}b). After 10 $\mu$s of expansion and before applying the magnetic gradient, we shine a radio-frequency pulse (of duration $30 ~\mu$s) to transfer a known fraction of $m_{J}=+1$ atoms into the $m_{J}=0$ state. 

\begin{figure}[ht!]
\includegraphics[width=\columnwidth]{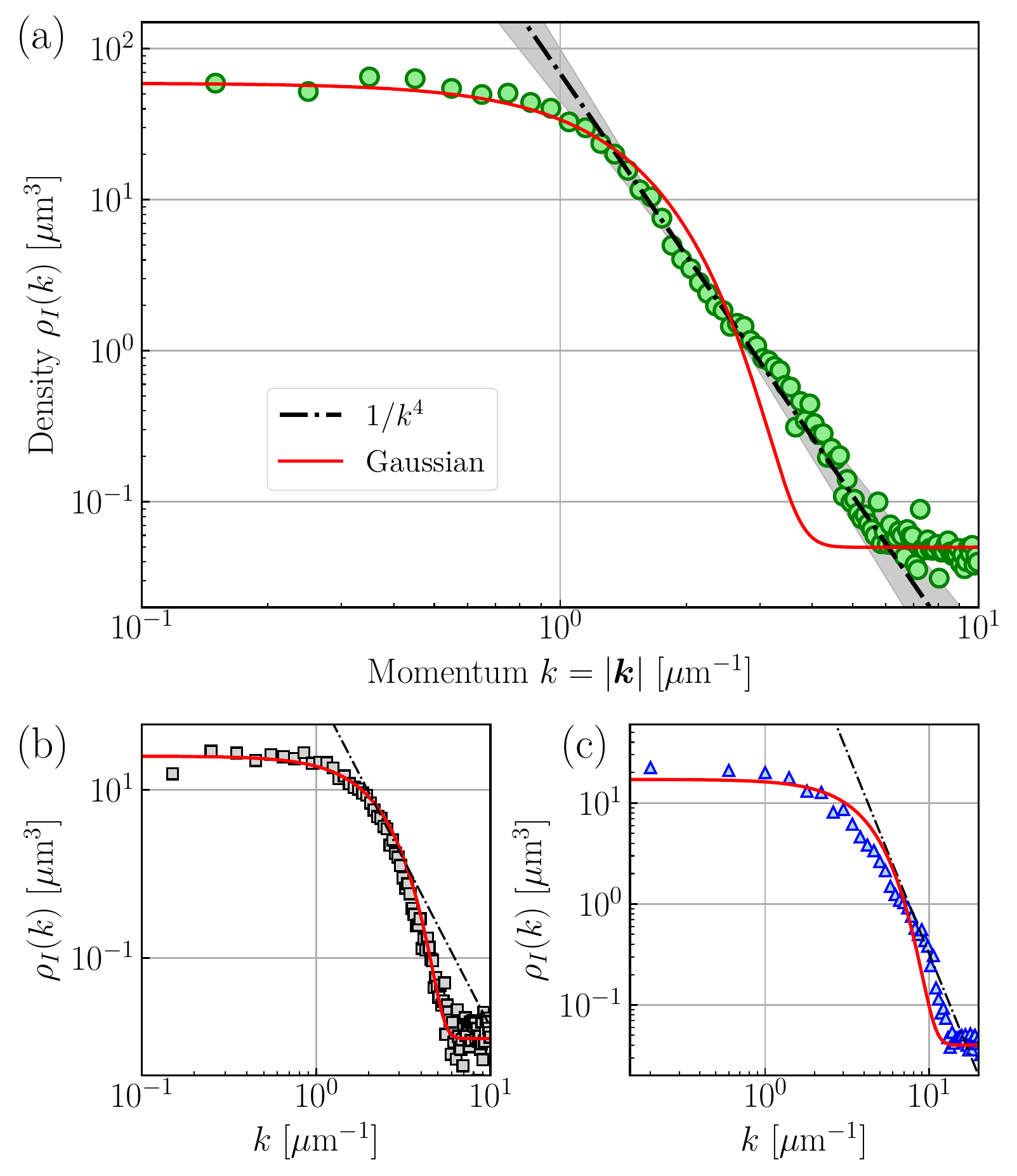}
\caption{
{\bf (a)} The momentum density $\rho_{\rm I}(k)$ of the impurity atoms in the presence of a spin-polarised BEC in $m_{J}=+1$ (green dots) exhibits algebraically decaying tails, consistent with $1/k^4$ (dashed-dotted black line). Power-law functions $1/k^{\alpha}$ with $3.5\leq \alpha \leq 4.5$ are also consistent with the data, as illustrated by the grey shaded area. The red (solid) line is a fit with a Gaussian function. {\bf (b)} $\rho_{\rm I}(k)$ in the absence of $m_{J}=+1$ atoms in the trap (black squares) is well-fitted by a Gaussian function corresponding to the momentum density of an ideal Bose gas at $T\simeq 200~$nK (red dashed line). 
{\bf (c)} $\rho_{\rm I}(k)$ in the presence of a thermal gas of $m_{J}=+1$ atoms in the trap (blue triangles). No algebraic decay is clearly identifiable.}
\label{fig2}
\end{figure}

In a first set of experiments, we measure the far-field momentum density $\rho_{\rm I}({\bm k})$ of the spin impurities ($m_{J}=0$), normalized so that $\int d^3{\bm k} \ \rho_{\rm I}({\bm k})=N_{I}$. The far-field density $\rho_{\rm I}({\bm k})$ measured after a long expansion differs from the in-trap momentum density as a result of the interactions which affect the early stages of the expansion. The 3D momentum ${\bm k}$ of an atom is determined from the measured 3D position ${\bm r}$ after time-of-flight using the ballistic relation, ${\bm k}=m {\bm r}/\hbar t_{\rm TOF}$. 
Importantly, in all the data sets presented in this work, $\rho_{\rm I}({\bm k})$ is found isotropic. This  allows us to increase dramatically the signal-to-noise ratio in $\rho_{\rm I}$ by taking a spherical average and study $\rho_{\rm I}$ as a function of the modulus $k=|{\bm k}|$, as shown in Fig.~\ref{fig2}. 
The dataset in Fig.~\ref{fig2}(a) is recorded with a number of atoms in the BEC $N_{\rm BEC}=4.4(3) \times 10^5$, and a fraction of impurities $f_{\rm I}\simeq 0.3(1) $\%. The asymptotic momentum density $\rho_{\rm I}$ exhibits algebraic tails whose decay is consistent with $1/k^4$. Note that the observed algebraic decay does not span a large enough dynamical range to precisely determine the power-law exponent \cite{ross2022}. Figs.~\ref{fig2}(b)-(c) show the same quantity measured in the absence of the BEC, in two different configurations. The density $\rho_{\rm I}$ shown in Fig.~\ref{fig2}(b) is recorded in the absence of $m_{J}=+1$ atoms, so that only the impurity cloud is present in the trap, and it is well fitted by a Gaussian modelling a thermal gas. To eliminate the BEC from the trap we added a magnetic gradient strong enough to expel the sub-level $m_{J} = +1$, while keeping the atoms in $m_{J} = 0$ optically trapped. In Fig.~\ref{fig2}(c), $\rho_{\rm I}$ is measured in the presence of a thermal gas of $m_J=+1$ atoms. In both panels Figs.~\ref{fig2}(b)-(c) no algebraic tails are identified, demonstrating that the algebraic tails are unambiguously observed in $\rho_{\rm I}$ only in the presence of a BEC.

\begin{figure}[h!]
\includegraphics[width=\columnwidth]{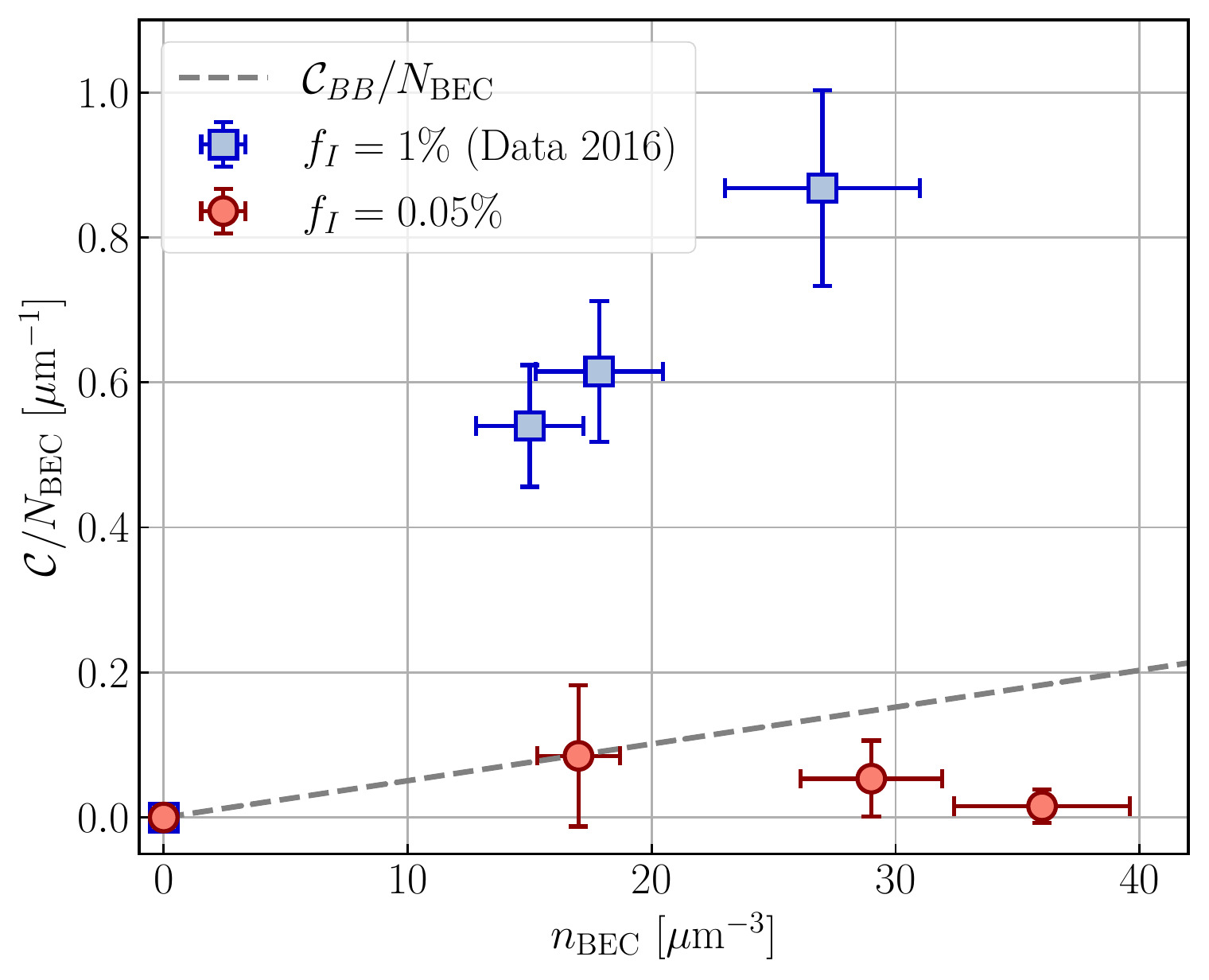}
\caption{Amplitude $\mathcal{C}$ of the $1/k^4$-tails in the BEC (normalised to $N_{\rm BEC}$) as a function of the BEC density, for two fractions of impurities $f_{\rm I}=1~$\% (blue squares) and $f_{\rm I}=0.05~$\% (red dots). The dashed-line is the in-trap contact $\mathcal{C}_{\rm Bogo}^{BB}/N_{\rm BEC}$ of a spin-polarised BEC predicted by Bogoliubov theory [see Eq.~(\ref{Eq:CBB})].
}
\label{Fig3}
\end{figure}

In a second set of experiments, we measure the asymptotic momentum density of a cloud in which a known fraction of the trapped $m_{J}=+1$ BEC atoms is transferred to the He$^*$ detector (see Fig.~\ref{fig1}b). 
We repeat the experiment and analysis performed in \cite{chang2016}, but here we reduce the fraction of spin impurities in the BEC to our smallest possible value $f_{\rm I} \simeq 0.05~$\%. For a direct comparison with \cite{chang2016}, we introduce the quantity $\mathcal{C}=(2 \pi)^3\mathcal{A}$ from the the measured amplitude $\mathcal{A}= k^4 \times \rho (k)$ of the tails (fitted over the momentum range $2 \mu $m$^{-1} - 6 \mu $m$^{-1}$), and in Fig.~\ref{Fig3} we plot $\mathcal{C}$ divided by the BEC atom number $N_{\rm BEC}$. We stress that $\mathcal{C}$ is obtained analysing the atomic densities measured after an expansion in the presence of interactions. Therefore, $\mathcal{C}$ is expected to differ from Tan's contact at equilibrium in the trap.

The values $\mathcal{C}/N_{\rm BEC}$ we measure with $f_{\rm I}=0.05\%$ lie much lower than those we found previously (where $f_{\rm I}$ was $1\%$) \cite{chang2016} and are consistent with zero (see Fig.~\ref{Fig3}). This uncovers the role played by the impurities in the findings we previously reported in \cite{chang2016}. In particular, this rules out the hypothesis that the tails are a direct signature of the quantum depletion of the spin-polarised BEC. Indeed, the resulting fitted amplitudes are much smaller than Tan's contact $\mathcal{C}_{\rm Bogo}^{BB}$ associated to the quantum depletion of a trapped spin-polarised BEC \cite{chang2016},
\begin{equation}
\mathcal{C}_{\rm Bogo}^{BB}=(64/7) \pi^2 a_{BB}^2 n_{\rm BEC} N_{\rm BEC}, \label{Eq:CBB}
\end{equation}
where $n_{\rm BEC}$ indicates the BEC density at the trap center. Eq.~(\ref{Eq:CBB}) is obtained starting from Tan's contact for a homogeneous BEC, $C_{0}=16 \pi^2 a_{BB}^2 n_{\rm BEC} N_{\rm BEC}$, and using a local density approximation (LDA) in the trap \cite{chang2016}. The LDA is safely applicable since $\xi \simeq 0.4~\mu$m $ \ll R_{z} \simeq 15~\mu$m, where $\xi$ is the BEC healing length and $R_{z}$ the smallest in-trap BEC radius. Our new observations confirm the scenario predicted theoretically in \cite{qu2016}, i.e., that the $1/k^4$-tails associated with the quantum depletion of a spin-polarised BEC decrease adiabatically during an expansion in the presence of interactions. The measurements and conclusions discussed here therefore differ from those of a recent work \cite{ross2022} which studied magnetically-trapped $^4$He$^*$ atoms. So far, unambiguous signals of the quantum depletion in time-of-flight experiments have been found only when interactions do not affect the expansion \cite{lopes2017, tenart2021}.

\begin{figure}[h!]
\includegraphics[width=\columnwidth]{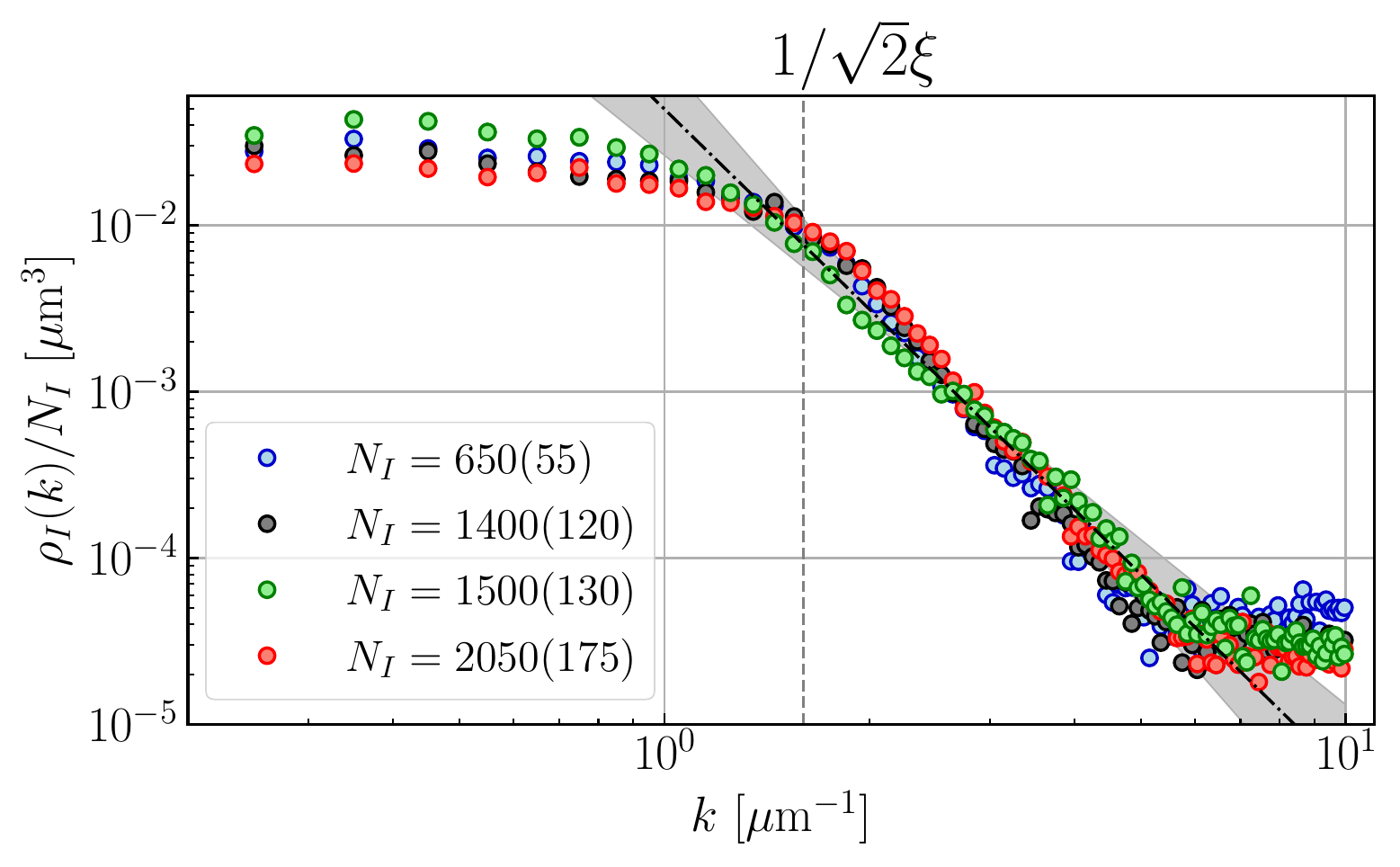}
\caption{Normalised momentum density $\rho_{I}(k)/N_{I}$ of impurities as a function of $k$. The data sets correspond to a varying impurity number $N_{I}$ in a BEC with a fixed total atom number $N_{\rm BEC}=4.5(5) \times 10^5$. The vertical dotted line indicates the momentum $k_{c, \rm BEC}=1/\sqrt{2}\xi$ of the sound velocity in the BEC. The dashed-dotted line (\emph{resp.} the grey shaded area) is a power-law function $1/k^{\alpha}$ with $\alpha=4$ (\emph{resp.} $3.5 \leq \alpha \leq 4.5$).
}
\label{fig6}
\end{figure}

\begin{figure}[ht!]
\includegraphics[width=\columnwidth]{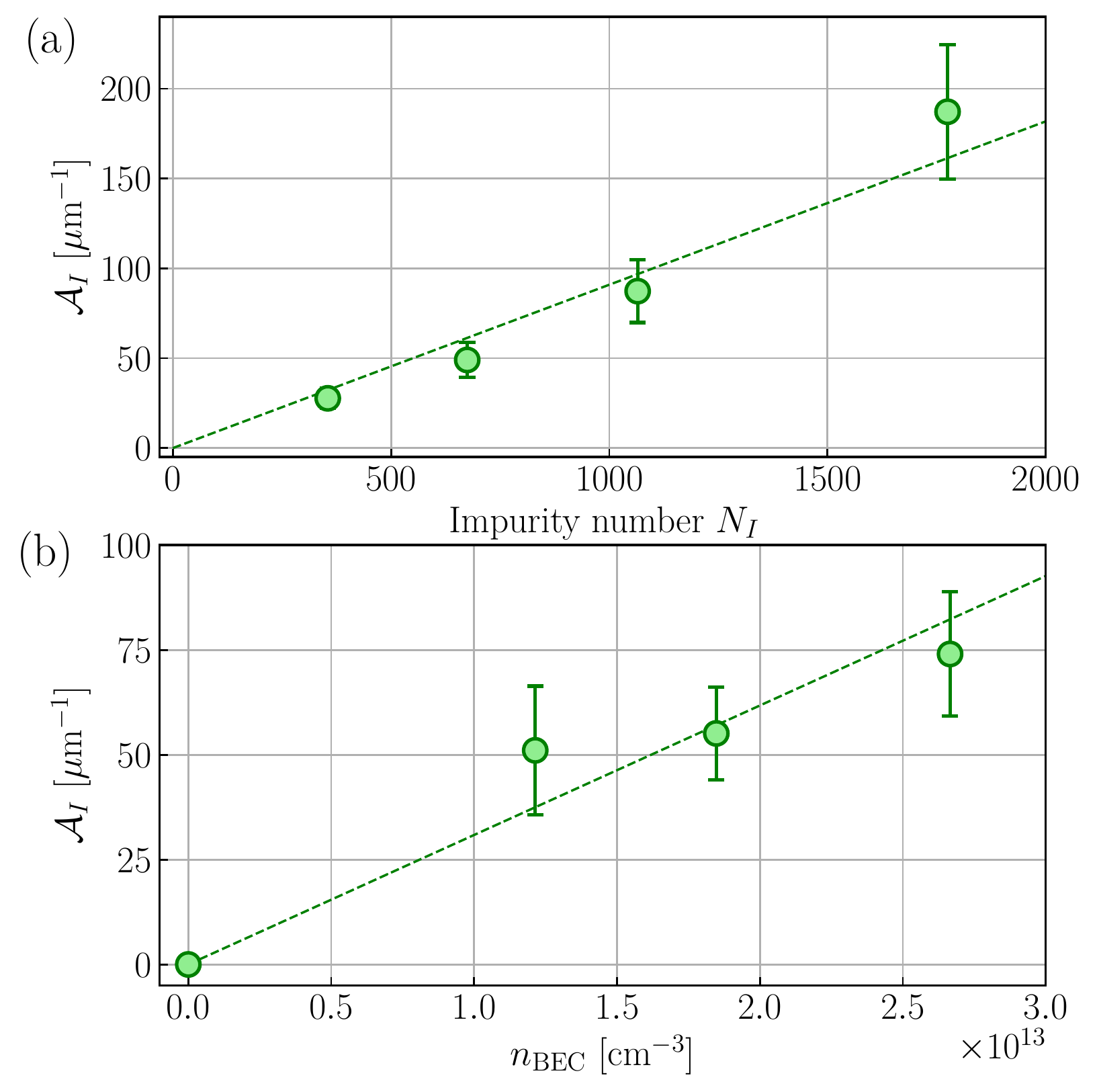}
\caption{ 
{\bf (a)} Amplitude $\mathcal{A}_{\rm I}$ of the $1/k^4$-tails in the impurity cloud as a function of impurity number $N_{I}$. The BEC atom number is fixed to $N_{\rm BEC}=5.5 \times 10^5$. {\bf (b)} Amplitude $\mathcal{A}_{\rm I}$ as a function of the BEC density $n_{\rm BEC}$ with a fixed $N_{I}=770$. The point at $n_{\rm BEC}=0$ corresponds to the data shown in Fig.~\ref{fig1}(b). 
In both panels, the green dashed line is a guide-to-the-eye.}
\label{fig4}
\end{figure}

\begin{figure}[ht!]
\includegraphics[width=\columnwidth]{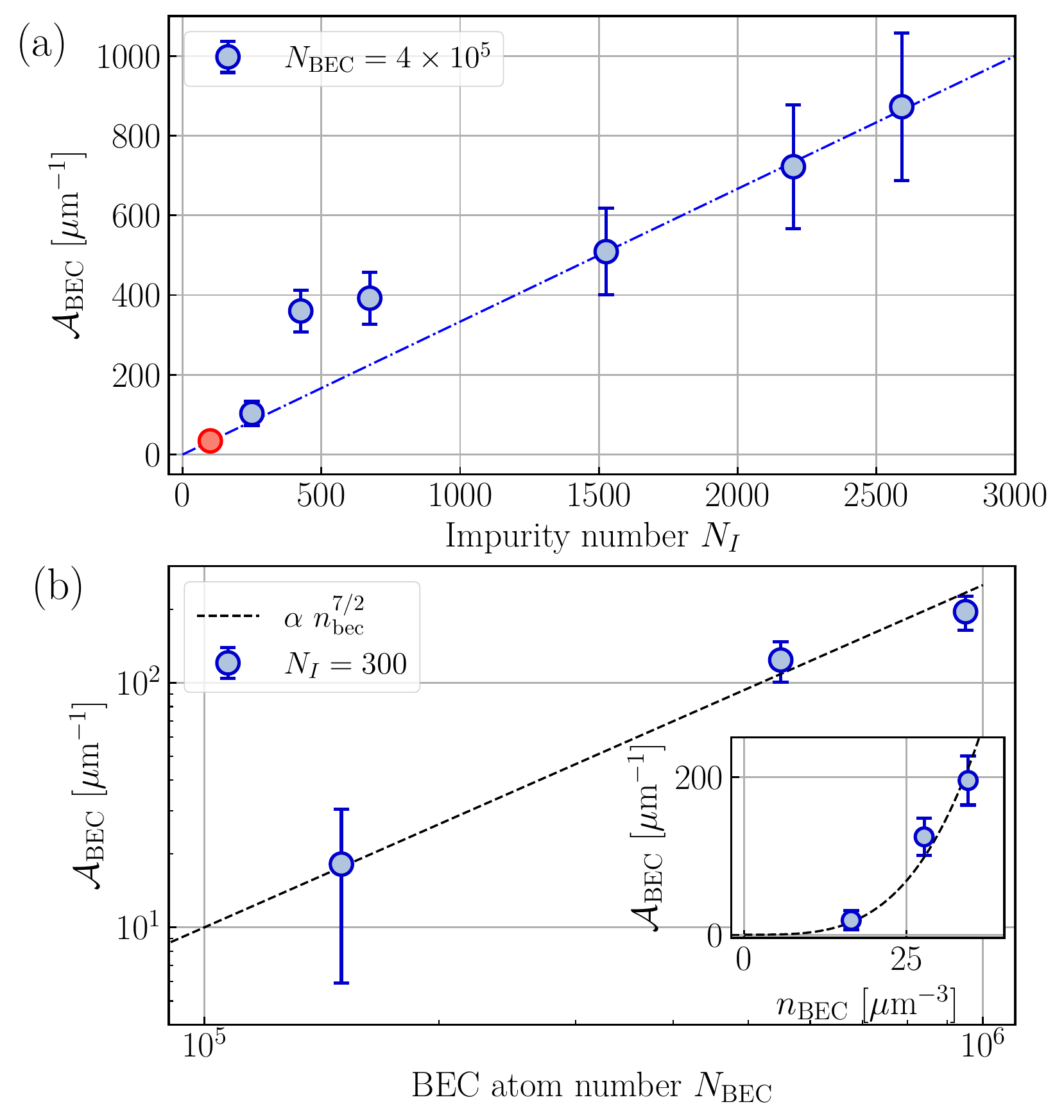}
\caption{
{\bf (a)} Amplitude $\mathcal{A}_{\rm BEC}$ of the $1/k^4$-tails in the BEC as a function of $N_{I}$. The red dot corresponds to the data set with $f_{I}\simeq 0.05 \%$ shown in Fig.~\ref{Fig3}. The blue dashed line is a guide-to-the-eye. {\bf (b)} Amplitude $\mathcal{A}_{\rm BEC}$ as a function of $N_{\rm BEC}$ with $N_{I}=300$. Inset: $\mathcal{A}_{\rm BEC}$ plotted as a function of the BEC density. In the main panel and the inset, the dashed black line is proportional to $n_{\rm BEC}^{7/2}$.}
\label{fig5}
\end{figure}

We turn to discussing the algebraic tails observed in the presence of both the BEC and impurities. To extract the tails amplitude $ \mathcal{A}_{\rm BEC}$ ({\it resp.} $\mathcal{A}_I$) in the bath ({\it resp.} the impurity cloud), we combine the analysis of the data recorded with the two detection methods shown in Fig.~\ref{fig1} \cite{NoteContact}. In a given data set, the tails amplitude $\mathcal{A}$ is obtained from fitting the plateau in $k^4 \times \rho(k)$ over the range $2 \mu $m$^{-1} - 6 \mu $m$^{-1}$. We have studied how $\mathcal{A}_{\rm BEC}$ and $\mathcal{A}_I$ change as we vary the number $N_{\rm I}$ of impurities and the BEC atom number $N_{\rm BEC}$. A series of such measurements in the impurity cloud is shown in Fig.~\ref{fig6} where $N_{I}$ is varied at fixed $N_{\rm BEC}$. The analysis of all the data sets is summarised in Figs.~\ref{fig4} and \ref{fig5} \cite{ChangeConfig}. 

In Fig.~\ref{fig4}(a), we plot the amplitude $\mathcal{A}_{I}$ in the impurity cloud as a function of the number of impurities. $\mathcal{A}_{I}$ increases to a good approximation linearly with $N_{I}$. 
In Fig.~\ref{fig4}(b), we observe that $\mathcal{A}_{I}$ also increases with the BEC density $n_{\rm BEC}$ at a fixed number of impurities. 
In Fig.~\ref{fig5}(a), the amplitude $\mathcal{A}_{\rm BEC}$ in the majority (BEC) atoms is plotted as a function of the number of impurities $N_{I}$. Its growth is consistent with a linear increase with $N_I$. Interestingly, the amplitude $\mathcal{A}_{\rm BEC}$ is found to vary rapidly with $n_{\rm BEC}$ at a fixed number of impurities [see Fig.~\ref{fig5}(b)], with a scaling $\propto n_{\rm BEC}^{7/2}$ similar to that of the bath-bath contact $\mathcal{C}_{\rm Bogo}^{BB}$ ($N_{\rm BEC}\propto n_{\rm BEC}^{5/2}$ in the Thomas-Fermi regime). 
We shall now discuss the possible origin of the algebraic tails.
\\

In the regime of weak interactions investigated in the experiment, the \emph{in-trap} momentum densities are known to have $1/k^4$-tails whose amplitudes are accurately described by Tan's contact. In the majority (BEC) atoms, Tan's contact results from the sum of two contributions: {\it (i)} the bath-bath interactions and {\it (ii)} the bath-impurity interactions. The bath-bath contact $\mathcal{C}_{\rm Bogo}^{BB}$ of Eq.~(\ref{Eq:CBB}) was derived in the LDA approximation. For the bath-impurity contact $\mathcal{C}_{\rm Bogo}^{BI}$ we also use the LDA approximation, assuming that the dilute impurities do not affect the BEC density profile in the trap. Since the impurities are trapped in a flat-bottom potential and the bath-impurity interaction is weak so that the lowest-order perturbative result holds, we find
\begin{equation}
\mathcal{C}_{\rm Bogo}^{BI}=(32/5) \pi^2 a_{BI}^2 n_{\rm BEC} N_{I}.
\label{Eq:Cbi}
\end{equation}
In the impurity cloud, the interaction between impurities is negligible due to their small concentration, and only $\mathcal{C}_{\rm Bogo}^{BI}$ plays a significant role in determining the in-trap amplitude of their momentum tails at large momenta. 
However, as discussed above for a BEC without impurities, these $1/k^4$ signatures of two-body contact interactions are expected to vanish during the time-of-flight expansion in the presence of interactions. The latter is indeed equivalent to probing the equation of state at progressively diminishing densities, with the consequence that interaction effects vanish after long expansions. This scenario does not match with our observations as the measured tails' amplitudes \emph{largely exceed} the \emph{in-trap} predictions. 
Therefore, our findings indicate that the algebraic tails result from the impurity-BEC interactions, but also that these are not a direct
manifestation of two-body interactions at equilibrium. We speculate that they result from the expansion dynamics, although we have not yet identified a scenario which explains our observations.

To complete our description, additional intriguing observations should be mentioned.  
First, the tails' amplitude in the BEC is comparable to that in the impurity cloud ($\mathcal{A}_{\rm BEC} \sim 2 - 3 \times \mathcal{A}_{I}$) while the BEC atom number exceeds that of impurities by several orders of magnitude ($N_{I} \lesssim N_{\rm BEC} /100$). This suggests that some equilibration between the momentum components of the impurities and of the BEC occurs during the expansion. 
Second, Fig.~\ref{fig6} shows that the algebraic tails are visible for momenta larger than $k_{c, \rm BEC}=1/\sqrt{2}\xi$, the momentum associated with the BEC sound velocity $c=\sqrt{g_{BB} n_{\rm BEC}/m}$. Impurities moving faster than the BEC sound velocity -- {\it i.e.}, the critical velocity for superfluidity -- are known to create excitations in the BEC \cite{astrakharchik2004}, contrary to slow moving impurities \cite{NoteThermalVsImpurity}. 
However, we performed GP simulations of the expansion dynamics of a BEC and a single impurity and we did not find $1/k^4$-tails in the momentum densities. Whether such effects play a role in our observations is an intriguing open question. 

In conclusion, we have studied the momentum densities of weakly-interacting Bose polarons formed when dilute spin impurities are immersed in a Bose-Einstein condensate and expand in the presence of interactions. The algebraic tails observed only in the presence of both the BEC and the impurities are not related to the in-trap contact, i.e.,  to short-range two-body physics at equilibrium, but rather result from the dynamics of moving impurities in an expanding superfluid bath. A detailed analysis of the complete expansion dynamics seems therefore crucial to understand our puzzling observations.
\\

{\it Acknowledgements.} We acknowledge fruitful discussions with C.~Carcy, M. Mancini, L. P. Pitaevskii, S. Stringari and all the members of the Quantum Gas group at Institut d'Optique, as well as technical help on the apparatus from G. Hercé. We acknowledge financial support from the R\'egion Ile-de-France in the framework of the DIM SIRTEQ, the ``Fondation d'entreprise iXcore pour la Recherche", the Agence Nationale pour la Recherche (Grant number ANR-17-CE30-0020-01). D.C.~acknowledges support from the Institut Universitaire de France. A.A.~acknowledges support from the Simons Foundation (Grant 601939) and from Nokia-Bell labs. P.M.~was supported by grant PID2020-113565GB-C21 funded by MCIN/AEI/10.13039/501100011033, by EU FEDER Quantumcat, by the National Science Foundation under Grant No. NSF PHY-1748958, and by the {\it ICREA Academia} program. This work was supported in part by the Swiss National Science Foundation under Division II.

\bibliography{MomentumTailsBib.bib}

\begin{thebibliography}{49}
\bibitem [{Exp()}]{ExpDetail}%
  \BibitemOpen
  \bibinfo {note} {To vary the BEC atom number, we add a magnetic gradient of
  varying amplitude which increases atom losses during the evaporation in the
  ODT; this procedure hardly changes the number of impurities because the
  latter are insensitive to the magnetic gradient. To vary the number of
  impurities, we hold the BEC and impurities in the ODT for holding times
  ranging from 0.1 to 3s. The value of the holding time strongly affects the
  impurity number while it hardly changes the BEC atom number.}\BibitemShut
  {Stop}%
\bibitem [{Not({\natexlab{a}})}]{NoteContact}%
  \BibitemOpen
  \bibinfo {note} {The data without RF transfer directly yield the tails
  amplitude $\mathcal{A}_{I}$ for the impurities only (up to a factor equal to
  the detection efficiency). To extract $\mathcal{A}_{\rm BEC}$ for the BEC
  atoms from the measured amplitude $\mathcal{A}$ with RF transfer, we proceed
  as follows. In the experimental configuration, the RF pulse transfers 34\%
  impurities outside of the detector and 16\% BEC atoms onto the detector.
  Because we have measured $\mathcal{A}_{I}$ in the absence of RF transfer, we
  subtract the contribution $0.66 \times \mathcal{A}_{I}$ from $\mathcal{A}$
  and extract the contribution $\mathcal{A}_{\rm BEC}$ from the BEC atoms only:
  $\mathcal{A}_{\rm BEC}=( \mathcal{A}-0.66 \times
  \mathcal{A}_{I})/0.16$.}\BibitemShut {Stop}%
\bibitem [{Cha()}]{ChangeConfig}%
  \BibitemOpen
  \bibinfo {note} {We have also performed additional experiments in slightly
  different configurations by changing the time-of-flight (from 298~ms to
  330~ms) and by applying the RF transfer after a 3~ms time-of-flight (instead
  of 10~$\mu$s). In all those configurations, we observe identical
  $1/k^4$-tails, which indicates that our observations are immune to these
  aspects of the experimental procedure.}\BibitemShut {Stop}%
\bibitem [{Not({\natexlab{b}})}]{NoteThermalVsImpurity}%
  \BibitemOpen
  \bibinfo {note} {Since we speculate that the algebraic tails may originate
  from scattering of fast impurities against the BEC, one could wonder if
  similar tails should appear when thermal atoms scatter against the BEC.
  However, we do not observe that in our experiment: when the fraction of
  impurities is negligible, thermal (majority) atoms are certainly present but
  $1/k^{4}$-tails are hardly discernable (see Fig.~3). This difference may stem
  from the fact that thermal (majority) atoms are located on the edge of the
  trap because they are repelled by the BEC, while the impurities are spread
  across the BEC as they are trapped in a flat-bottom potential. Fast
  impurities (above the sound velocity) have therefore a much larger chance to
  cross the BEC and scatter than the thermal (majority) atoms.}\BibitemShut
  {Stop}%
\end{thebibliography}

\end{document}